\documentclass[%
 amsmath,amssymb,
twocolumn, superscriptaddress
]{revtex4-1}

\usepackage[normalem]{ulem}
\usepackage{cancel}

\usepackage[dvipsnames]{xcolor}
\usepackage{graphicx}
\usepackage{dcolumn}
\usepackage{bm}
\usepackage{orcidlink}
\hypersetup{colorlinks, citecolor=BrickRed}
\hypersetup{colorlinks=true, linkcolor=Blue, urlcolor=MidnightBlue}
\usepackage{hyperref}
\usepackage{ulem}
\usepackage{xcolor}
\usepackage{footnote}

 \newcommand{\be}{\begin{equation}}
\newcommand{\ee}{\end{equation}}
\newcommand{\bea}{\begin{eqnarray}}
\newcommand{\eea}{\end{eqnarray}}

\begin{document}

\preprint{APS/123-QED}

\title{Universal relations for anisotropic interacting quark stars}

\author{Juan M. Z. Pretel}
 \email{juanzarate@cbpf.br}
 \affiliation{
 Centro Brasileiro de Pesquisas F{\'i}sicas, Rua Dr.~Xavier Sigaud, 150 URCA, Rio de Janeiro CEP 22290-180, RJ, Brazil
}

\author{Chen Zhang}
 \email{iasczhang@ust.hk (Corresponding author)}
 \affiliation{
 The HKUST Jockey Club Institute for Advanced Study, The Hong Kong University of Science and Technology, Hong Kong S.A.R., P.R.China
}
\date{\today}
\begin{abstract}
Interacting quark stars, which are entirely composed of interacting quark matter including perturbative QCD corrections and color superconductivity, can meet constraints from various pulsar observations. In realistic scenarios, pressure anisotropies are expected in the star's interior. Recently, the stellar structural properties of anisotropic interacting quark stars have been investigated. In this study, we further explore the universal relations (URs) related to the moment of inertia $I$, tidal deformability $\Lambda$, compactness $C$, and the $f$-mode nonradial pulsation frequency for such stars. Our results reveal that these approximate URs generally hold, being insensitive to both the EOS variations as well as to the presence of anisotropy. In contrast to previous studies on anisotropic neutron stars, we find that more positive anisotropy tends to enhance the $I-\Lambda$ and $I-C$ URs, but weakens the $C-\Lambda$ UR. For all the URs involving $f$-mode frequency, we find that they are enhanced by the inclusion of anisotropy (whether positive or negative).  Utilizing these URs and the tidal deformability constraint from the GW170817 event, we put limits on the structural properties of isotropic and anisotropic quark stars, such as the moment of inertia $I_{1.4}$, the canonical radius $R_{1.4}$ and the canonical $f$-mode frequency $f_{f,1.4}$, all of which are very different compared to those of neutron stars.
\end{abstract}

\maketitle


\section{Introduction} 
Empirically, it is known that hadronic matter (HM) is bound at nuclear densities to form neutron stars (NSs). However, it is theoretically possible that compact stars are described by quark degrees of freedom, a matter phase called quark matter (QM), which can form bare quark stars (QSs) that are highly compact. 

 The Bodmer-Witten-Terazawa hypothesis~\cite{Bodmer:1971we, Witten:1984rs, Terazawa:1979hq} suggests that QM with comparable amounts of $u, \,d, \,s$ quarks, also called strange quark matter (SQM), might be the ground state of baryonic matter at low (zero) temperature and pressure. A recent study~\cite{Holdom:2017gdc} demonstrated that $u, d$ quark matter ($ud$QM) is, in general, more stable than SQM and the ordinary nuclear matter at a sufficiently large baryon number beyond the periodic table. Such bulk absolute stability allows the possibility of up-down quark stars consisting of $ud$QM~\cite{Zhang:2019mqb, Ren:2020tll, Wang:2019jze,Xia:2020byy, Xia:2022tvx, Cao:2020zxi, Yuan:2022dxb, Wang:2021byk, Li:2022vof, Restrepo:2022wqn}, in addition to strange quark stars consisting of SQM~\cite{Bodmer:1971we, Witten:1984rs, Terazawa:1979hq, Farhi:1984qu, Alford:2007xm, Weber:2004kj, Haensel:1986qb, Alcock:1986hz, Zhou:2017pha, Weissenborn:2011qu, Xu:1999bw, Yang:2023haz}. Interacting quark matter (IQM) includes the interquark effects induced by strong interaction, such as the perturbative QCD (pQCD) corrections~\cite{Farhi:1984qu, Fraga:2001id, Fraga:2013qra} and the color superconductivity~\cite{Alford:1998mk, Rajagopal:2000ff, Lugones:2002va}. Recently, it was shown that the equation of state (EOS) of IQM can be recast into a simple unified form and the resulting interacting quark stars (IQSs) can reconcile various astrophysical constraints~\cite{Zhang:2020jmb}, which has motivated a series of follow-up studies~\cite{Zhang:2021iah,Zhang:2021fla,Blaschke:2021poc,Blaschke:2022egm,Oikonomou:2023otn,Gammon:2023uss,Tangphati:2023fey,Yuan:2023dxl,Zhou:2024syq,Gammon:2024gij} using this IQM EOS. In a realistic scenario, anisotropy effects should also be accounted for in QSs, which have been an active research area ~\cite{Komathiraj:2007cw, Harko:2002pxr, Kalam:2012wb, Sunzu:2014wva, Panahi:2015vza, Deb:2016lvi, Deb:2018ccw, Arbanil:2016wud, Arbanil:2023yil, Das:2023qej, Lopes:2023phz}. Very recently, anisotropic effects on the stellar structure of IQSs have been systematically studied~\cite{Pretel:2023nlr}.

Anisotropy in compact stars can naturally arise from various factors, such as large magnetic fields, phase transitions, relativistic nuclear interactions, crystallization, or superfluidity of the core. Several phenomenological models have been proposed to incorporate anisotropic pressure in compact stars, such as the Bowers-Liang (BL) model~\cite{BowersLiang1974}, Quasi-Local (QL) profile~\cite{Horvat2011}, Herrera and Barreto (HB) model~\cite{HerreraBarreto2013}, etc. In general, the anisotropy may affect various macroscopic properties, such as mass, radius, moment of inertia, tidal deformability~\cite{Pretel:2023nlr}, rotational properties~\cite{Pattersons:2021lci,Beltracchi2024}, radial~\cite{Pretel:2020xuo, Mohanty2024} and nonradial oscillations~\cite{Arbanil:2023yil, Mohanty:2023hha}. The existence of pressure anisotropy can help explain some recent observations~\cite{biswas2019tidal,Das:2023qej,becerra2024realistic}, and can lift the star compactness significantly so that anisotropic stars could feature gravitational-wave echoes in the late stage of their formation~\cite{raposo2019anisotropic}. However, there is a degeneracy on comparable effects from the EOS variations. It is thus physically important to investigate the effect of anisotropy on universal relations (URs) that are EOS insensitive.

The various URs between these quantities are insensitive to EOSs, and thus can be used to extract information from one to the others that may not be easy observables, breaking degeneracies that would otherwise limit our ability to take full advantage of pulsar observations~\cite{YagiYunes2013}. The URs for the moment of inertia ($I$), tidal deformability (Love), and spin-induced quadrupole moment ($Q$) were first derived by Yagi and Yunes for NSs that are slowly rotating, tidally deformed, isotropic~\cite{Yagi:2013bca,YagiYunes2013} or anisotropic~\cite{Yagi2015}, and generally hold both for conventional NSs and QSs~\cite{YagiYunes2017,Chan2016,Sham:2014kea,Yeung:2021wvt}, except for some violations found in hybrid stars~\cite{Annala:2017tqz,Han:2018mtj,10.1063/1.5117807,Hoyos:2021uff} and in stars with two fluids differentially rotating~\cite{Yeung:2021wvt,Cronin:2023xzc}. There are also other URs that involve star compactness ($C$)~\cite{Maselli:2013mva,Chan2015,Chan2016}, and fundamental mode frequency ($f$) of non-radial oscillations~\cite{Andersson1998, Lau2010, ChanSLL2014, Chirenti2015, ZhaoLattimer2022}. In particular, the $f$-mode can have imprints in the gravitational-wave signals of the binary merger events~\cite{hinderer2016effects, pratten2020gravitational,miao2024resolving,yu2024dynamical} from its coupling to dynamical tides. Recently, $I-\text{Love}-C$ and $I-f-C$ URs for anisotropic NSs were studied in Ref.~\cite{Das:2022ell} using the BL model and Ref.~\cite{Mohanty:2023hha} using the QL model, and it was found that anisotropy will generally weaken the universal relations in BL model for the former, but will enhance or weaken the URs in QL model depending on the sign of anisotropy parameters for the latter. In this study, we generalize the exploration of these URs to anisotropic quark stars. This is highly nontrivial since it is known that for isotropic cases, the URs related to compactness are quite different between NSs and QSs~\cite{
YagiYunes2017,Maselli:2013mva, Chan2016}. As we will show below, anisotropic quark stars also have quite distinct URs involving compactness compared to those of anisotropic neutron stars for given anisotropy.  Correspondingly, the derived limits on structural quantities like the canonical radius inferred from the Love-$C$ UR and the tidal deformability bound of LIGO/Virgo binary events will also be very different.

As for the organization of this paper, we first review the details of IQM. We then introduce the QL anisotropy profile, where a free parameter $\beta_{\rm QL}$ controls the amount of anisotropy within the IQS. After that, we apply this anisotropy model to the stellar properties of IQSs, such as the mass-radius, moment of inertia, tidal deformability, and nonradial oscillations under the Cowling approximation, with the main focus on their URs. All stellar structure equations will be rewritten in a dimensionless rescaled form so that the different macroscopic properties and the URs do not depend on the specific value of the effective bag constant $B_{\rm eff}$. Finally, we discuss the sensitivity of these universal relations to the quark matter EOSs and anisotropy parameter variations. Throughout this work, we adopt the geometrized unit system where $G = 1 = c$.


\section{Modeling of anisotropic interacting quark stars} 

\subsection{Equation of state for radial pressure}
\label{IQM_EOS}
Our study involves anisotropic stellar fluids where in addition to a radial pressure $p_r$ there is a transverse pressure $p_t$. Here we will begin by describing the equation of state for radial pressure. Referring to~\cite{Alford:2004pf, Zhang:2020jmb}, we first rewrite the thermodynamic potential $\Omega$ of interacting quark matter with the superconducting effect~\cite{Alford:2002kj} and the pQCD correction included:  
\begin{align}
  \Omega=&-\frac{\xi_4}{4\pi^2}\mu^4+\frac{\xi_4(1-a_4)}{4\pi^2}\mu^4- \frac{ \xi_{2a} \Delta^2-\xi_{2b} m_s^2}{\pi^2}  \mu^2  \nonumber  \\
  &-\frac{\mu_{e}^4}{12 \pi^2}+B_{\rm eff} ,
\label{omega_mu}
\end{align}
where $\mu$ and $\mu_e$ are the respective average quark and electron chemical potentials. The first term represents the unpaired free quark gas contribution. The second term with $(1-a_4)$ represents the pQCD contribution from one-gluon exchange for gluon interaction to $O(\alpha_s^2)$ order. To phenomenologically account for  higher-order contributions, we can vary $a_4$ from $a_4=1$, corresponding to a vanishing pQCD correction, to very small values where QCD corrections become large~\cite{Fraga:2001id, Alford:2004pf, Weissenborn:2011qu}. The term with $m_s$ accounts for the correction from the finite strange quark mass if applicable. The term with the gap parameter $\Delta$ represents the contribution from color superconductivity.  $(\xi_4,\xi_{2a}, \xi_{2b})$ represents different state of color-superconducting phases. Moreover, $B_{\rm eff}$ denotes the effective bag constant, accounting for the nonperturbative contribution from QCD vacuum.

The corresponding EOS for the radial pressure was derived in Ref.~\cite{Zhang:2020jmb}:
\begin{align}
  p_r =&\ \frac{1}{3}(\rho-4B_{\rm eff})  \nonumber \\
  &+ \frac{4\lambda^2}{9\pi^2}\left[ -1+ {\rm sgn}(\lambda)\sqrt{1+3\pi^2 \frac{(\rho-B_{\rm eff})}{\lambda^2}}\right],  \label{eos_tot}
\end{align}
where \be
\lambda=\frac{\xi_{2a} \Delta^2-\xi_{2b} m_s^2}{\sqrt{\xi_4 a_4}}.
\label{lam}
\ee 
Note that $\rm sgn(\lambda)$ represents the sign of $\lambda$. For this study, we need to explore only positive $\lambda$ space, as preferred by astrophysical observations~\cite{Zhang:2020jmb}. This simple form of interacting quark matter model not only covers but also generalizes the parameter space studied by various renowned literature, such as~\cite{Zhou:2017pha,Alford:2004pf,Weissenborn:2011qu}.

As shown in Ref.~\cite{Zhang:2020jmb}, one can further remove the $B_{\rm eff}$ parameter by doing the following dimensionless rescaling:
\be
\tilde{\rho}=\frac{\rho}{4B_{\rm eff}}, \qquad  \tilde{p}_r=\frac{p_r}{4B_{\rm eff}}, 
\label{scaling_prho}
\ee
and
\be
 \tilde{\lambda}=\frac{\lambda^2}{4B_{\rm eff}}= \frac{(\xi_{2a} \Delta^2-\xi_{2b} m_s^2)^2}{4B_{\rm eff}\xi_4 a_4},
 \label{scaling_lam}
\ee
so that the EOS~(\ref{eos_tot}) reduces to the dimensionless form
\be
\tilde{p}_r =\frac{1}{3}(\tilde{\rho}-1)+ \frac{4}{9\pi^2}\tilde{\lambda} \left[-1+\sqrt{1+\frac{3\pi^2}{\tilde{\lambda}} {\left(\tilde{\rho}-\frac{1}{4}\right)}}\right].
\label{eos_p}
\ee

As $\tilde{\lambda}\to0$, Eq.~(\ref{eos_p}) reduces to the rescaled conventional noninteracting quark matter EOS  $\tilde{p}_r =(\tilde{\rho}-1)/3$. On the other hand, when $\tilde{\lambda}$ becomes extremely large, Eq.~(\ref{eos_p}) approaches the special form
\be
\tilde{p}_r\vert_{\tilde{\lambda}\to \infty}= \tilde{\rho}-\frac{1}{2}, 
\label{eos_infty}
\ee
or, equivalently, $p_r={\rho}-2B_{\rm eff}$, using Eq.~(\ref{scaling_prho}). We see that strong interaction effects can reduce the surface mass density of a quark star from $\rho_0= 4B_{\rm eff}$ down to $\rho_0=2B_{\rm eff}$ and increase the quark matter radial sound speed $c_{rs}^2 =\partial p_r/\partial \rho$ from $1/3$ up to $1$ (the light speed) maximally.

\subsection{Anisotropy profile}\label{AnisotropyProfile}

Anisotropic pressure in compact star systems may be a manifestation of several factors such as strong magnetic fields, phase transitions, pion condensation, rotation, bosonic composition, etc., see the review papers \cite{HerreraSantos1997, Kumar2022} for further details. Consequently, anisotropic fluid configurations have been intensively analyzed in recent years using various ansatzes. The literature offers some well-motivated functional relations for anisotropy, which must satisfy general physical requirements or ``physical acceptability conditions'' \cite{Mak2003, Abreu2007}:~1) Energy density, radial pressure and tangential pressure should be positive anywhere within the anisotropic star, 2) gradients for energy density and radial pressure must be negative, 3) the radial and tangential speed of sound should be less than the speed of light, 4) the interior solution must satisfy the strong and dominant energy conditions, 5) regularity at the stellar center, among others. In our analysis, we will adopt the phenomenological model that allows the construction of anisotropic stars obeying the aforementioned conditions. Originally introduced by Horvat and collaborators \cite{Horvat2011}, the Quasi-Local ansatz is given by
\begin{equation}\label{QLProfileEq}
    \sigma = \beta_{\rm QL}\left( \frac{2m(r)}{r} \right)p_r ,
\end{equation}
where $\beta_{\rm QL}$ is a dimensionless constant that measures the degree of anisotropy within the relativistic fluid sphere and $m(r)$ is the mass function. In this study, we adopt the range of $\beta_{\rm QL}$ within the interval $[-0.6, 0.6]$ for which appreciable changes in the mass-radius diagrams can be visualized while being consistent with observations~\cite{Pretel:2023nlr,Curi2022} and other studies~\cite{Mohanty:2023hha}.

The above anisotropy profile has two important traits: First, it allows regularity at the stellar origin since the fluid becomes isotropic there, and second, because it arises only within a relativistic regime, i.e., its contribution is zero in the Newtonian limit. This last characteristic is of physical importance because it has been argued that anisotropy may arises only at very high energy densities \cite{Yagi2015, Folomeev2018}, and other profiles such as the Bowers-Liang model \cite{BowersLiang1974} do not satisfy this condition. The astrophysical implications of Eq.~(\ref{QLProfileEq}) on compact stars have been intensively investigated both in conventional Einstein gravity \cite{Doneva2012, Yagi2015, Pretel:2020xuo, Rahmansyah2020, Rahmansyah2021, Arbanil:2023yil, Pretel:2023nlr, Curi2022} and in modified gravity theories \cite{Silva2015, Folomeev2018, Pretel2022, Pretel2022MPL, Tangphati2023, Li2023PDU, Pretel2024PS}.

\subsection{Mass-radius diagram}

In Einstein gravity, the most basic properties of a compact star such as mass and radius are the results of numerically integrating the Tolman-Oppenheimer-Volkoff (TOV) equations~\cite{Oppenheimer:1939ne,Tolman:1939jz}:
\begin{align}
  \frac{dm}{dr} &= 4\pi r^2 \rho , \label{tov1}\\
  \frac{dp_r}{dr} &= (\rho+ p_r)\, \frac{m+ 4\pi r^3 p_r }{r\left(2m- r\right)} + \frac{2\sigma}{r}, \label{tov2}
\end{align}
which are the product of considering an anisotropic perfect fluid described by the following energy-momentum tensor
\begin{equation}\label{EqEMTensor}
    T_{\mu\nu}= (\rho+ p_t)u_\mu u_\nu + p_t g_{\mu\nu} - \sigma k_\mu k_\nu , 
\end{equation}
with $\rho$, $p_r$ and $p_t$ being the energy density, radial pressure and tangential pressure, respectively. Here, $\sigma= p_t- p_r$ is the anisotropic factor, $k^\mu$ is a unit spacelike four-vector and $u^\mu$ is the four-velocity of the fluid. The spacetime geometry of the stellar system is described by the usual line element
\be
  ds^2=-e^{2\Phi(r)}dt^2+ e^{2\Psi(r)}dr^2+ r^2(d\theta^2+\sin^2\theta d\phi^2), \label{metric}
\ee
where $\Phi(r)$ and $\Psi(r)$ are the metric functions to be determined. The gravitational mass enclosed within a sphere of radius $r$, denoted by $m(r)$, is related to the metric function $\Psi(r)$ by means of $e^{-2\Psi(r)} = 1- 2m(r)/r$. The surface radius $R$ is calculated at which the radial pressure vanishes (i.e., by the condition $p_r(R)=0$), and the total mass of the star is $M= m(R)$.

Meanwhile, the metric potential $\Phi(r)$ is obtained after solving the differential equation \cite{Pretel:2020xuo}
\be
  \frac{d\Phi}{dr}= -\frac{1}{\rho+ p_r}\frac{dp_r}{dr} + \frac{2\sigma}{r(\rho+ p_r)} , \label{tov3}
\ee
whose solution will be useful later when calculating other macroscopic properties of anisotropic IQSs. The boundary condition for (\ref{tov3}) arises from the continuity of the interior metric (\ref{metric}) with the Schwarzschild exterior solution at the surface, namely
\be\label{surCondition}
  e^{2\Phi}= e^{-2\Psi}= 1-\frac{2M}{R}.
\ee

Following the dimensionless rescaling (\ref{scaling_prho}), we have the rescaled variables for the anisotropy factor, radial coordinate and mass function as
\begin{align}
    \tilde{\sigma} &=\frac{\sigma}{4B_{\rm eff}},  &  \tilde{r} &= r\sqrt{4B_{\rm eff}},  &  \tilde{m} &= m\sqrt{4B_{\rm eff}} ,
\end{align}
so the surface radius and total mass of the star will be denoted by $\tilde{R}$ and $\tilde{M}$, respectively. The new dimensionless variables will be taken into account in all stellar structure equations so that the calculations are independent of the value of $B_{\rm eff}$. 

\begin{figure*}
 \includegraphics[width=5.95cm]{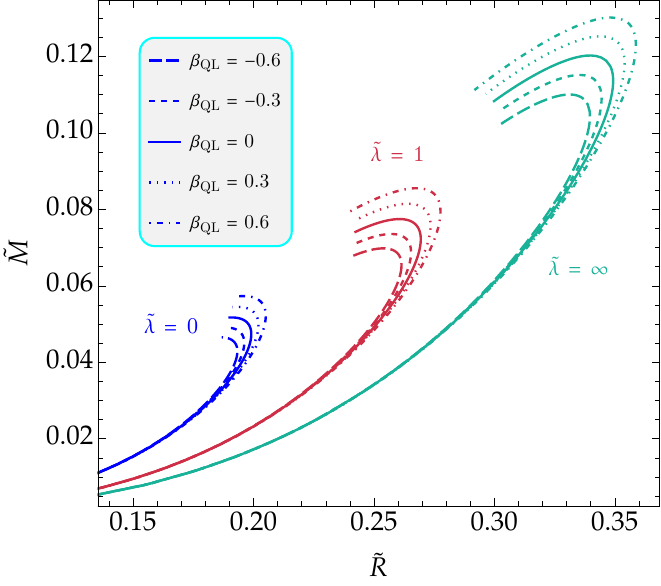}
 \includegraphics[width=5.758cm]{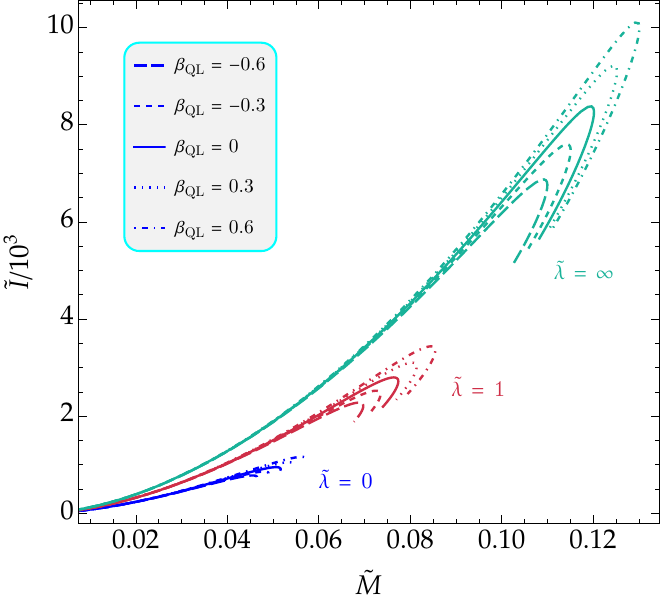}
  \includegraphics[width=5.965cm]{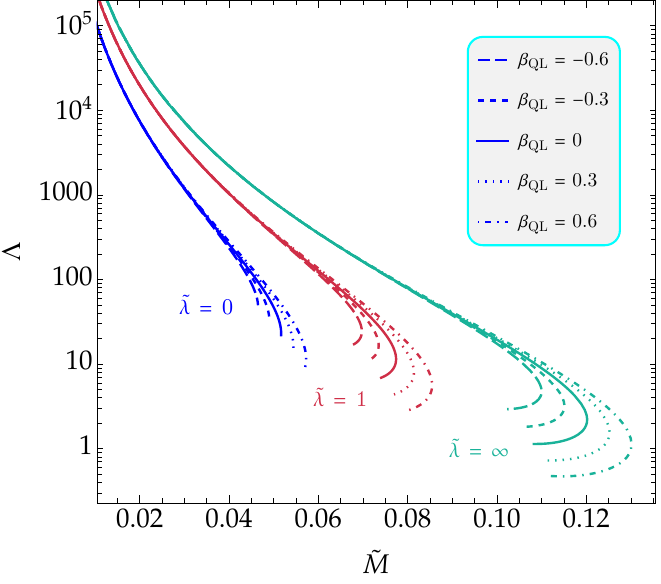}
 \caption{\label{fig_MR_I_L} From left to right, (a) Rescaled mass vs rescaled radius, (b) Dimensionless moment of inertia vs rescaled mass, and (c) Tidal deformability vs rescaled mass  for anisotropic IQSs with $\tilde{\lambda}= 0$, $1$ and $\infty$ for the interacting quark matter EOS (\ref{eos_p}) with QL anisotropy ansatz (\ref{QLProfileEq}), where we have considered $\beta_{\rm QL} \in [-0.6, 0.6]$. }
\end{figure*}

We begin our analysis by generating the rescaled mass-radius relations for several values of $\tilde{\lambda}$ and $\beta_{\rm QL}$. To this purpose, we numerically solve the system of equations (\ref{tov1}) and (\ref{tov2}) along with Eq.~(\ref{QLProfileEq}) by using the initial conditions $\rho(0)= \rho_c$ and $m(0)= 0$. Given a range of central densities, Fig.~\ref{fig_MR_I_L}a displays the mass versus radius for anisotropic IQSs, where we have used five specific values of $\beta_{\rm QL}$ and the solid curves stand for the isotropic solutions when $\beta_{\rm QL}= 0$. For the adopted anisotropy profile we have also considered three benchmark values of $\tilde{\lambda}= 0,1\, {\rm and}\, \infty$, where $\tilde{\lambda}= 0$ represents the standard noninteracting quark matter EOS. Remark that a larger $\tilde{\lambda}$ leads to more massive quark stars. It can be observed that positive (negative) anisotropies give rise to a substantial increase (decrease) in maximum-mass values of IQSs.

\section{Moment of inertia and tidal deformability} 

Under the anisotropic context and in the slowly rotating approximation \cite{Hartle1967}, the relativistic moment of inertia is calculated by first solving the following second-order differential equation \cite{Pretel:2023nlr}
\be\label{OmegaEq}
    \frac{e^{\Phi-\Psi}}{r^4}\frac{d}{dr}\left[ e^{-(\Phi+\Psi)}r^4\frac{d\varpi}{dr} \right] = 16\pi(\rho+ p_r+ \sigma)\varpi ,
\ee
for the frame-dragging function $\varpi(r)$. This expression is integrated from the origin at $r=0$ for an arbitrary choice of $\varpi(0)$ and with vanishing slope, i.e., $d\varpi/dr= 0$. Besides, the asymptotic flatness requirement has to be satisfied at very far distances from the stellar surface, so that $\lim_{r\rightarrow\infty}\varpi = \Omega$, where $\Omega$ is the angular velocity of the star. Thus, once the dragging angular velocity $\varpi(r)$ is found, the moment of inertia is determined via the integral
\begin{equation}\label{MomInerEq}
    I = \frac{8\pi}{3}\int_0^R (\rho+ p_r)\left[ 1+ \frac{\sigma}{\rho+ p_r} \right]\frac{r^4\varpi}{\Omega}  e^{\Psi-\Phi} dr .
\end{equation}

It is important to emphasize that the frame-dragging function and moment of inertia in their dimensionless rescaled forms are given by
\begin{align}
    \tilde{\varpi} &= \frac{\varpi}{\sqrt{4B_{\rm eff}}},  &  \tilde{I} &= I\sqrt{(4B_{\rm eff})^3} .
\end{align}

For each value of central energy density $\rho_c$ and given the anisotropy model (\ref{QLProfileEq}), Eq.~(\ref{OmegaEq}) along with the TOV equations are simultaneously solved. Our results for the integral (\ref{MomInerEq}) are illustrated in Fig.~\ref{fig_MR_I_L}b, where we present the dimensionless moment of inertia $\tilde{I}$ as a function of the total gravitational mass $\tilde{M}$. As the strong interaction effects become increasingly larger (i.e., as $\tilde{\lambda}$ increases), the moment of inertia is subjected to a relevant increase with respect to the noninteracting case. We further observe that the maximum values of $\tilde{I}$ increase as $\beta_{\rm QL}$ increases from negative values.

The tidal perturbation $y(r)$ for the spacetime metric of the anisotropic star is governed by the equation \cite{Arbanil:2023yil}
\begin{equation}\label{yEq}
    ry' = -y^2 + (1 - r\mathcal{A})y - r^2\mathcal{B} ,
\end{equation}
where
\begin{align}
    \mathcal{A} =&\ \frac{2}{r} + e^{2\Psi}\left[ \frac{2m}{r^2} + 4\pi r(p_r - \rho) \right] ,  \\
    \mathcal{B} =&\ 4\pi e^{2\Psi}\left[ 4\rho + 4p_r+ 4p_t + \frac{\rho+ p_r}{\mathcal{F}v_{sr}^2}(1+ v_{sr}^2) \right]  \nonumber  \\
    &- \frac{6e^{2\Psi}}{r^2} - 4\Phi'^2 ,
\end{align}
with $\mathcal{F} = dp_t/dp_r$ and $v_{sr}^2= dp_r/d\rho$. Through the initial condition $y(0)=2$ \cite{Postnikov2010} at the center, we determine the surface tidal perturbation $y(R)$ by integrating the differential equation (\ref{yEq}) from the origin up to the surface of the anisotropic interacting quark star. The dimensionless tidal deformability is given by $\Lambda = 2k_2/3C^5$, where $C= M/R$ is the compactness and tidal Love number $k_2$ has the form \cite{Hinderer2010, Arbanil:2023yil}:
\begin{align}\label{LoveNumEq}
    k_2 =&\ \frac{8}{5}(1- 2C)^2C^5 \left[ 2C(\alpha -1) - \alpha+ 2 \right]  \nonumber  \\
    &\times \left\lbrace 2C[ 4(\alpha+ 1)C^4 + (6\alpha- 4)C^3 \right.  \nonumber  \\
    &\left.+\ (26- 22\alpha)C^2 + 3(5\alpha -8)C - 3\alpha+ 6 \right]   \nonumber  \\
    &\left.+\ 3(1-2C)^2\left[ 2C(\alpha- 1)- \alpha +2 \right]\ln(1-2C) \right\rbrace^{-1} .
\end{align}

Note that, for compact stars with a nonzero density at the surface, $\alpha= y(R)- 4\pi R^3 \rho_s/M$ in the last expression, where $\rho_s$ is the density just inside the stellar surface. Furthermore, the rescaling quantities for the new variables introduced take the form
\begin{align}
    \tilde{\mathcal{A}} &= \frac{\mathcal{A}}{\sqrt{4B_{\rm eff}}} ,  &  \tilde{\mathcal{B}} &= \frac{\mathcal{B}}{4B_{\rm eff}} .
\end{align}

Equation (\ref{yEq}) must be integrated self-consistently with
the TOV equations (\ref{tov1}), (\ref{tov2}) and (\ref{tov3}) from the center of the star, where $\rho(0)= \rho_c$, to its surface at $r=R$. Fig.~\ref{fig_MR_I_L}c exhibits the dimensionless tidal deformability as a function of total mass for the adopted anisotropy model. We observe that $\Lambda$ becomes large for anisotropic IQSs near the small-mass configuration. Given a fixed value of $\tilde{M}$ in the small-mass region, an increase in $\tilde{\lambda}$ leads to obtaining larger tidal deformabilities with respect to the noninteracting case.

\begin{figure*}
 \includegraphics[width=17.8cm]{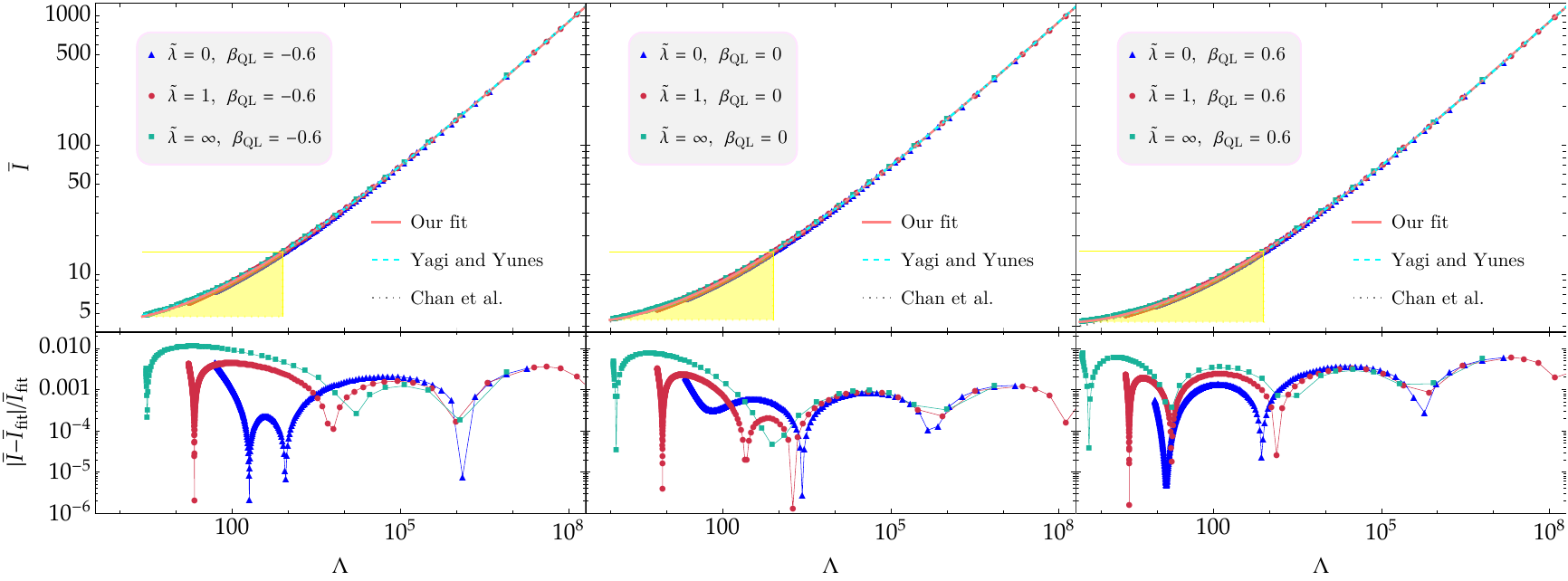}
 \includegraphics[width=17.8cm]{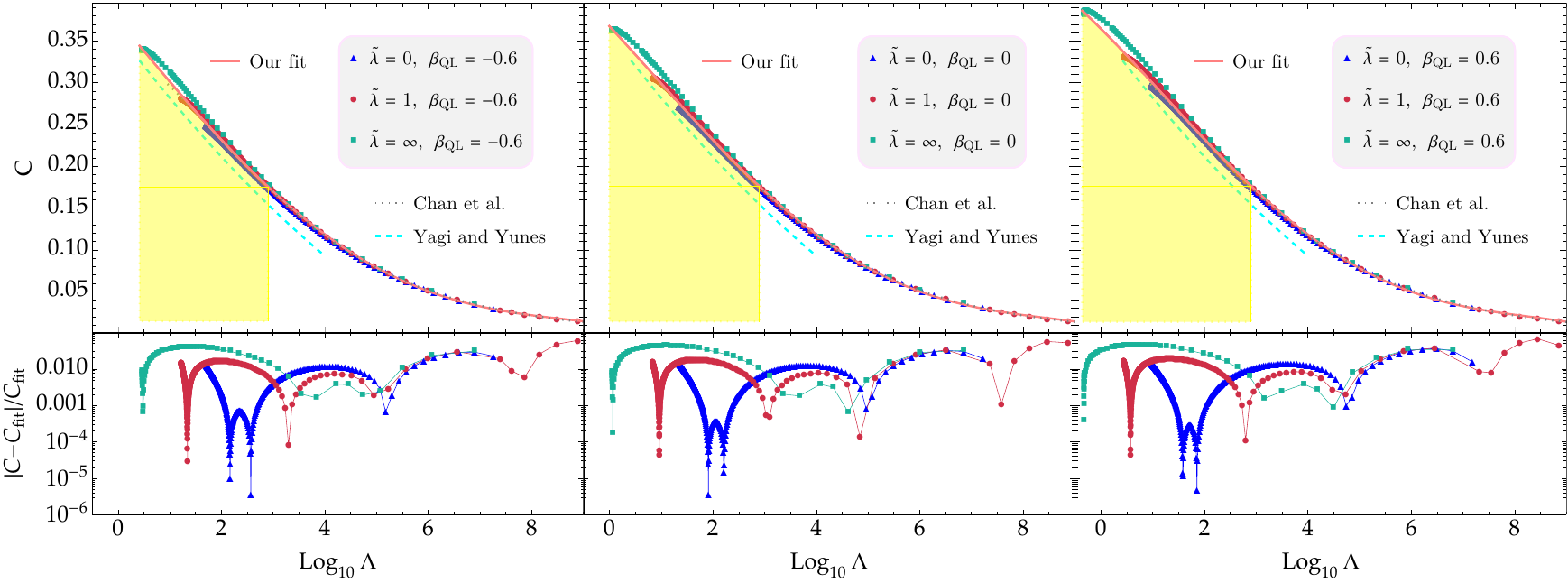}
 \includegraphics[width=17.8cm]{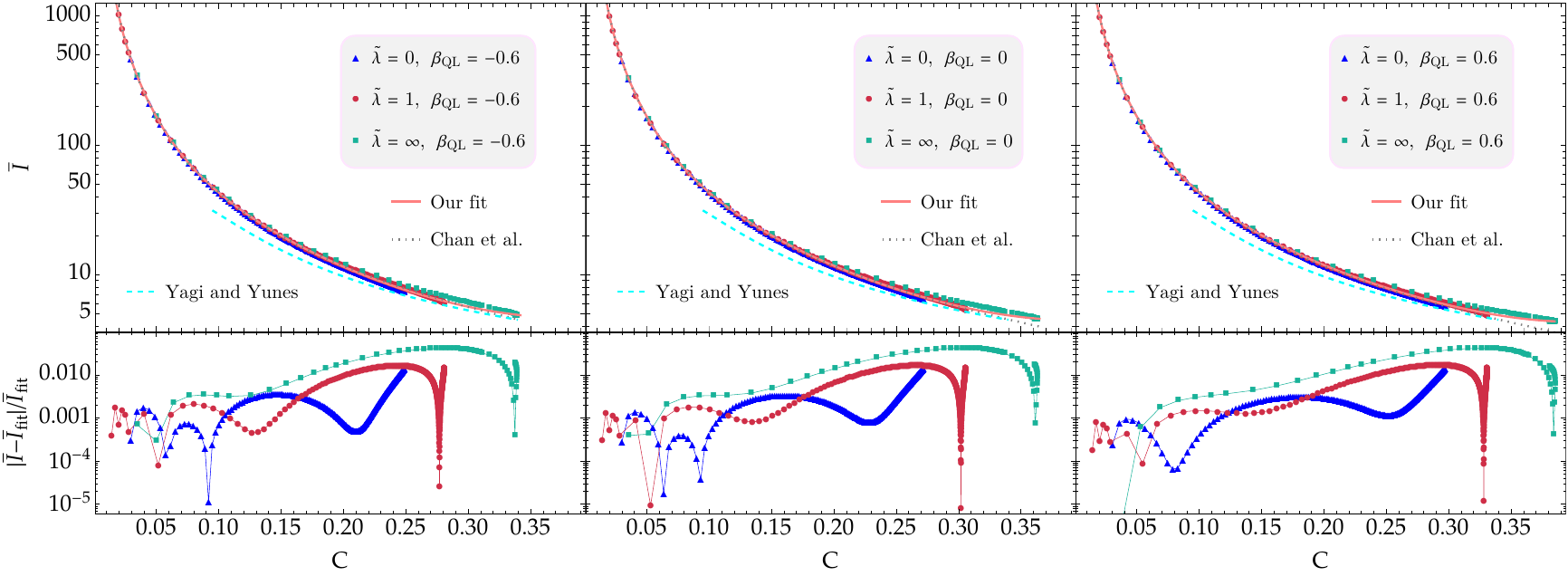}
 \caption{\label{figUnivRelationsQL} From top to bottom, (a) $\bar{I}-\Lambda$, (b) $C-\Lambda$ and (c) $\bar{I}-C$ universal relations for anisotropic IQSs with the QL anisotropy ansatz (\ref{QLProfileEq}), where we have assumed three benchmark values of $\beta_{\rm QL}= -0.6$ (left column), $0$ (middle column corresponding to the isotropic scenario) and $0.6$ (right column). Different symbols indicate the respective EOS for the radial pressure with three benchmark values of $\tilde{\lambda}$. Pink solid lines represent our fitted curves using the series expansions (\ref{UREqILambda}), (\ref{UREqCLambda}) and (\ref{UREqIC}) of corresponding URs, where the lower panel in each plot displays the relative fractional errors between fits and the numerical results. The filled yellow region stands for the EOS-independent constraint $\Lambda_{1.4} \leq 800$ from the GW170817 signal \cite{Abbott2017PRL}. Additionally, the gray dotted lines are the fitting functions for QSs derived within the post-Minkowskian approximation \cite{Chan2016}, while the cyan dashed curves are the analytical expressions for NSs \cite{YagiYunes2017}.  }  
\end{figure*}

It has been shown that the global properties of a compact star (such as radius, mass, moment of inertia and tidal deformability) are connected by universal relations \cite{YagiYunes2013, Chan2015, Chan2016, Breu2016, YagiYunes2017}. Under this perspective, in the present work we intend to analyze these universalities in the context of IQSs under the presence of anisotropic pressure. We begin by analyzing the $I-\Lambda$ relation through the normalized moment of inertia $\bar{I}= \tilde{I}/\tilde{M}^3$ \cite{Breu2016}. Given a specific anisotropy model, $\bar{I}$ is plotted as a function of the dimensionless tidal deformability $\Lambda$, see Fig.~\ref{figUnivRelationsQL}a for the QL ansatz where we have considered three benchmark values of $\beta_{\rm QL}$. We fit these relations with the following power series expansion
\be\label{UREqILambda}
  \log_{10}\bar{I} = \sum_{n=0}^4 a_n \left( \log_{10}\Lambda \right)^n ,
\ee
where $a_n$ are fitting coefficients with their respective reduced chi-squared ($\chi_{\rm red}^2$) values listed in Table \ref{table1}.  We can note that an increase in anisotropy (e.g., $\beta_{\rm QL}= 0.6$) gives rise to a decrease in $\chi_{\rm red}^2$ value with respect to the isotropic case, which indicates that the EOS-insensitive UR of $\bar{I}-\Lambda$ gets enhanced with the addition of positive anisotropy, while the opposite occurs for negative anisotropies.

\begin{figure*}
 \includegraphics[width=5.933cm]{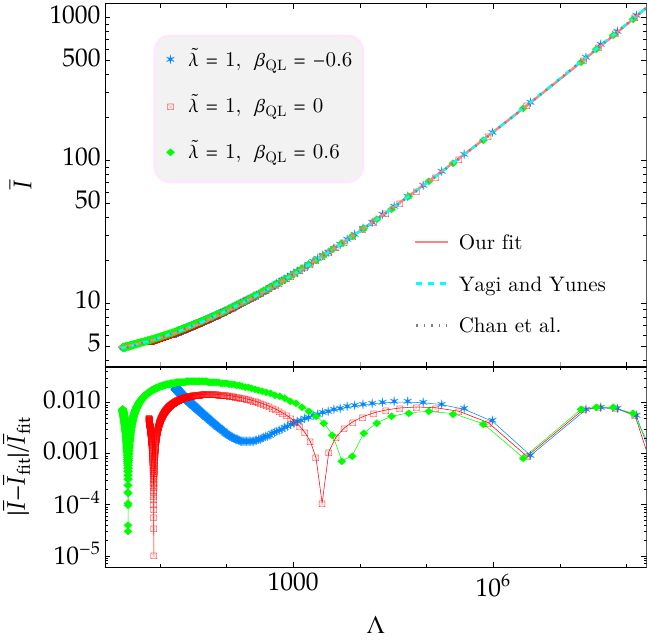}
 \includegraphics[width=5.895cm]{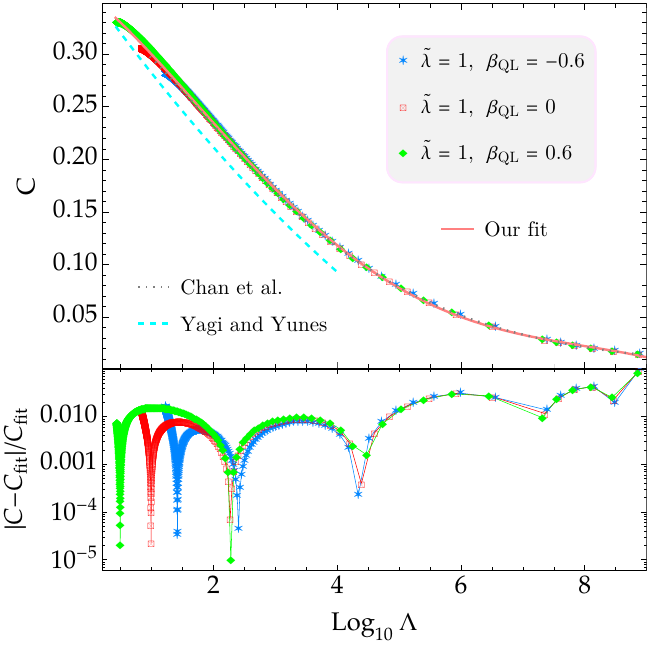}
 \includegraphics[width=5.895cm]{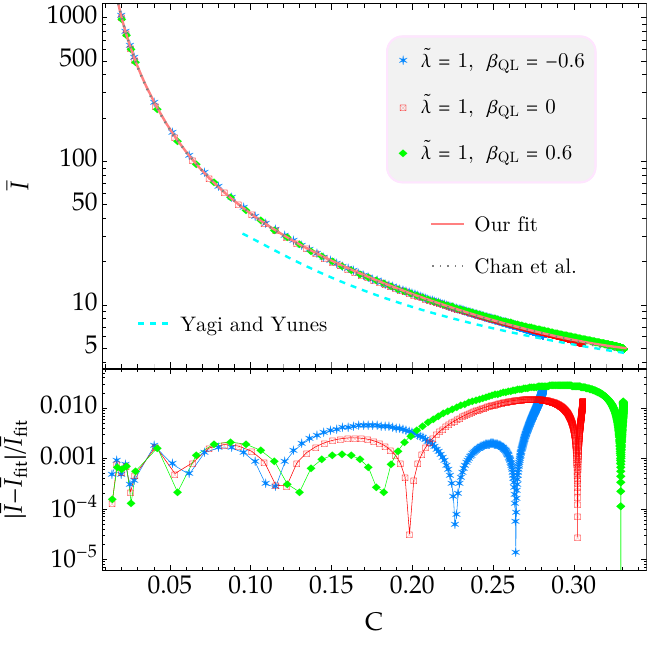}
    \caption{From left to right, (a) $\bar{I}-\Lambda$, (b) $C-\Lambda$ and (c) $\bar{I}-C$ URs for anisotropic IQSs of a benchmark EOS choice $\tilde{\lambda}= 1$, with the QL anisotropy parameter varying in the range $\vert\beta_{\rm QL}\vert \leq 0.6$. Pink solid lines represent our fitted curves, with the lower panel in each plot showing the relative fractional errors between fits and the numerical results. According to the relative fractional errors, these correlations remain approximately universal to $3\%$, $10\%$ and $4\%$ under variations of $\beta_{\rm QL}$, respectively. The results for other $\tilde{\lambda}$ choices share similar features.}
    \label{fig_UR_varyb} 
\end{figure*}

The lower panels in Fig.~\ref{figUnivRelationsQL}a display the absolute fractional difference between all numerical data (extracted by considering different EOSs with $\tilde{\lambda}= 0, 1, \infty$) and the fitting expression (\ref{UREqILambda}). It can be observed that, given a particular value of $\beta_{\rm QL}$, the $\bar{I}-\Lambda$ relation remains approximately universal with maximum
deviations at $\sim1\%$ level in the range of adopted $\tilde{\lambda}$, which is constrained by different observational measurements \cite{Zhang:2020jmb, Pretel:2023nlr}. Thus, this relation is insensitive to the EOS of interacting quark stars even if there is anisotropic pressure (see the left and right plots for $\beta_{\rm QL} \neq 0$) within the stellar fluid. Through a post-Minkowskian expansion for incompressible stars, Chan et al.~\cite{Chan2015} found analytic expressions for $\bar{I}$ and $\Lambda$ to sixth-order in compactness. A year later, the same authors extended this analysis to self-bound stars, which include quark stars, and they showed that the $\bar{I}-\Lambda$ relation is very similar to that for incompressible stars \cite{Chan2016}. In Fig.~\ref{figUnivRelationsQL}a, we have included the fit obtained by Chan and collaborators \cite{Chan2016} for the case of QSs by a gray dotted line. For comparison, we have also included the fitted curve for NSs obtained by Yagi and Yunes \cite{YagiYunes2017} by a cyan dashed line. We observe that the $\bar{I}-\Lambda$ correlation of NSs is almost indistinguishable from that of IQSs. We can therefore say that our $\bar{I}-\rm Love$ results are consistent with the analyses for other dense-matter EOSs. Furthermore, in Fig.~\ref{fig_UR_varyb}a we provide the $\bar{I}-\Lambda$ UR when the anisotropy parameter varies in the range $\vert\beta_{\rm QL}\vert \leq 0.6$ but $\tilde{\lambda}$ remains fixed to a benchmark value $\tilde{\lambda}=1$. In this case, the $\bar{I}-\Lambda$ relation remains approximately universal with maximum deviations at the $3\%$ level.

\begin{table}
\begin{minipage}[b]{80mm}
\caption{\label{table1} Fitting coefficients of the $\bar{I}-\Lambda$ UR given by the series expansion (\ref{UREqILambda}), with the anisotropy model (\ref{QLProfileEq}) for three benchmark values of the anisotropy parameter $\beta_{\rm QL}$. 
Moreover, the reduced chi-squared $(\chi_{\rm red}^2)$ values are shown for all cases in the last row. }
\begin{ruledtabular}
\begin{tabular}{c|ccc}
$\beta_{\rm QL}$  &  $-0.6$  &  0  &  0.6  \\
\hline
  $a_0 [10^{-1}]$  &  6.4232  &  6.4749  &  6.5096  \\
  $a_1 [10^{-2}]$  &  5.2637  &  6.2737  &  7.0025  \\
  $a_2 [10^{-2}]$  &  5.6645  &  5.0556  &  4.5931  \\
  $a_3 [10^{-3}]$  &  $-4.6588$  &  $-3.5984$  &  $-2.7527$   \\
  $a_4 [10^{-4}]$  &  1.5807  &  1.0003  &  0.5178  \\
  $\chi_{\rm red}^2 [10^{-6}]$  &  1.7645  &  0.5813  &  0.5612 
\end{tabular}
\end{ruledtabular}
\end{minipage}

\begin{minipage}[b]{80mm}
\caption{\label{table2} Fitting coefficients of the $C-\Lambda$ UR given by the series expansion (\ref{UREqCLambda}), with the same values of $\beta_{\rm QL}$ as in Table \ref{table1}. 
The respective values of $\chi_{\rm red}^2$ are given in the last row. }
\begin{ruledtabular}
\begin{tabular}{c|ccc}
$\beta_{\rm QL}$  &  $-0.6$  &  0  &  0.6  \\
\hline
  $b_0 [10^{-1}]$  &  3.7539  &  3.6766  &  3.6448  \\
  $b_1 [10^{-2}]$  &  $-7.4469$  &  $-6.6742$  &  $-6.1748$   \\
  $b_2 [10^{-3}]$  &  $-0.3753$  &  $-2.9372$  &  $-4.9372$   \\
  $b_3 [10^{-3}]$  &  0.9488  &  1.2898  &  1.5839  \\
  $b_4 [10^{-5}]$  &  $-5.3656$  &  $-6.9389$  &  $-8.3793$   \\
  $\chi_{\rm red}^2 [10^{-5}]$  &  0.7736  &  1.0498  &  1.3769
\end{tabular}
\end{ruledtabular}
\end{minipage}

\begin{minipage}[b]{80mm}
\caption{\label{table3} Fitting parameters of the $\bar{I}-C$ UR given by the series expansion (\ref{UREqIC}), with the same values of $\beta_{\rm QL}$ as in Table \ref{table1}. The respective values of $\chi_{\rm red}^2$ are given in the last row. }
\begin{ruledtabular}
\begin{tabular}{c|ccc}
$\beta_{\rm QL}$  &  $-0.6$  &  0  &  0.6  \\
\hline
  $c_{-4} [10^{-7}]$  &  $-1.3945$  &  $-1.0597$  &  $-0.6584$  \\
  $c_{-3} [10^{-5}]$  &  2.8396  &  2.1490  &  1.3560   \\
  $c_{-2} [10^{-3}]$  &  $-2.2886$  &  $-1.7453$  &  $-1.1512$   \\
  $c_{-1} [10^{-2}]$  &  0.1036  &  8.2858  &  6.1387   \\
  $c_{0}$  &  0.8531  &  1.2671  &  1.6668   \\
  $c_{1}$  &  0.9939  &  $-3.3393$  &  $-7.2093$   \\
  $c_{2} [10^{1}]$  &  $-1.9843$  &  $0.3404$  &  $2.2466$   \\
  $c_{3} [10^{1}]$  &  $4.8042$  &  $-0.9847$  &  $-5.2987$   \\
  $c_{4} [10^{1}]$  &  $-2.8670$  &  $2.3648$  &  $5.8645$   \\
  $\chi_{\rm red}^2 [10^{-5}]$  &  7.8028  &  6.9737  &  5.9877
\end{tabular}
\end{ruledtabular}
\end{minipage}
\end{table}

\begin{table}
\caption{\label{tableFromGW170817} Theoretical limits for the canonical moment of inertia, radius and fundamental mode frequency from the tidal deformability bound given by the GW170817 event \cite{Abbott2017PRL} for a $1.4M_\odot$ compact star. The different regions limited by this bound are represented by yellow areas in Figs.~\ref{figUnivRelationsQL} and \ref{figOLUnivRelationQL}. } 
\begin{ruledtabular}
\begin{tabular}{c|ccc}
$\beta_{\rm QL}$  &  $I\, [10^{45}\, \rm g\cdot cm^2]$  &  $R\, [\rm km]$  &  $f_f= \omega_f/2\pi\, [\rm kHz]$  \\
\hline
  $-0.6$  &  $I_{1.4} \leq 1.763$  &  $R_{1.4} \leq 11.793$  &  $f_{f,1.4} \geq 2.004$  \\
  $0$  &  $I_{1.4} \leq 1.783$  &  $R_{1.4} \leq 11.770$  &  $f_{f,1.4} \geq 2.108$  \\
  $0.6$  &  $I_{1.4} \leq 1.795$  &  $R_{1.4} \leq 11.728$  &  $f_{f,1.4} \geq 2.213$   
\end{tabular}
\end{ruledtabular}
\end{table}

Fig.~\ref{figUnivRelationsQL}b displays the $C-\rm Love$ UR for the three EOSs, where each plot from left to right corresponds to a fixed value of $\beta_{\rm QL}$ as in Fig.~\ref{figUnivRelationsQL}a. A fit function of the form
\be\label{UREqCLambda}
  C = \sum_{n=0}^4 b_n \left( \log_{10}\Lambda \right)^n
\ee
is also included in each plot by a pink curve. The best-fit coefficients are given in Table \ref{table2}. Contrary to what happened with the $\bar{I}-\Lambda$ correlation, here we see that positive anisotropies increase the value of $\chi_{\rm red}^2$ which weakens the $C-\Lambda$ EOS independent relation. Nevertheless, negative anisotropies make this UR stronger. We have included the fits obtained for NSs \cite{YagiYunes2017} and QSs \cite{Chan2016} by cyan dashed and gray dotted lines, respectively. Our results reveal that the $C-\Lambda$ correlation of neutron stars is appreciably distinguishable from that of quark stars. However, our findings are compatible with the fit obtained by Chan et al.~\cite{Chan2016}. According to the bottom panels, the fractional difference between the data and the fit (\ref{UREqCLambda}) show that the $C-\Lambda$ relation is approximately universal with maximum
deviations at $\sim 8\%$ level, which is much larger than that of the $\bar{I}-\Lambda$ UR in Fig.~\ref{figUnivRelationsQL}a, whose maximum equation-of-state variation led to fractional errors of the order of $1\%$. From Fig.~\ref{fig_UR_varyb}b, we observe also that this relation is insensitive to the variation of $\beta_{\rm QL}$ to $\sim 10\%$ level for fixed $\tilde{\lambda}= 1$.

Fig.~\ref{figUnivRelationsQL}c exhibits the relation between $\bar{I}$ and $C$, for which our fit (pink curve) is built based on the following polynomial function
\be\label{UREqIC}
  \log_{10}\bar{I} = \sum_{n=-4}^4 c_n C^n ,
\ee
where the fitting parameters $c_n$ with their corresponding chi-squared ($\chi_{\rm red}^2$) errors are presented in Table \ref{table3}. Our findings reveal that positive (negative) values of $\beta_{\rm QL}$ lead to a stronger (weaker) EOS insensitive $\bar{I}-C$ relation. Furthermore, according to the lower panels in Fig.~\ref{figUnivRelationsQL}c, the maximum fractional difference for the compact stars sequence is at least $5\%$, less than that of the $C-\Lambda$ UR but still much greater than that obtained in the $\bar{I}-\Lambda$ correlation. Remarkably, the $\bar{I}-C$ relation of IQSs is quite different from that of neutron-star matter (see the cyan dashed curve). Nonetheless, our fitting formula is consistent with the analytic expression for quark stars derived by Chan et al.~\cite{Chan2016}, as shown by the gray dotted line. Moreover, Fig.~\ref{fig_UR_varyb}c shows the $\bar{I}-C$ UR for fixed $\tilde{\lambda}= 1$, which is insensitive to variations of $\beta_{\rm QL}$ with maximum deviations at the $4\%$ level. Therefore, the different URs analyzed so far remain insensitive not only to variations in the EOS of IQM but also to the presence of anisotropy (when $\beta_{\rm QL} \neq 0$).

Using our URs, tidal deformability data can be very useful in imposing constraints on the canonical radius and moment of inertia, namely, two important macroscopic properties for a $1.4M_\odot$ compact star. The tidal deformability bound $\Lambda_{1.4}= 190_{-120}^{+390}$, reported by LIGO-Virgo Collaboration from the GW170817 signal \cite{Abbott2018PRL}, is based on the assumption that the coalescing bodies were NSs. Nonetheless, since the present study is dealing with quark stars, we must use the EOS-independent constraint $\Lambda_{1.4} \leq 800$ from the older paper \cite{Abbott2017PRL}. Such a bound has been included in all correlations involving $\Lambda$, see yellow regions in the first two rows of Fig.~\ref{figUnivRelationsQL}. For instance, for the isotropic case (i.e., when $\beta_{\rm QL} = 0$), the $\bar{I}-\Lambda$ UR allows us to obtain $\bar{I}_{1.4} \leq 14.954$. Taking into account that $\bar{I} = c^4I/G^2M^3$ in physical units, this limit can be rewritten as $I_{1.4} \leq 1.783 \times 10^{45}\, \rm g\cdot cm^2$. Similarly, from the $C-\Lambda$ relation and the above tidal deformability restriction, we can obtain an upper limit for the radius $R_{1.4}$. Considering three benchmark values of $\beta_{\rm QL}$, the theoretical limits for $I_{1.4}$ and $R_{1.4}$ can be found in Table \ref{tableFromGW170817}.

It is worth commenting that in the case of NSs, the limit on the canonical radius differs significantly from our result for IQSs due to the appreciable difference in the $C-\Lambda$ relation obtained for hadronic matter (see the cyan dashed curve) and quark matter (the pink solid line for our fit). This difference is attributed to the fact that they are stellar systems of different microscopic composition. The canonical radius $R_{1.4}$ from the GW170817 event was obtained by Das \cite{Das:2022ell} using URs for anisotropic NSs, where he considered that the anisotropic pressure is described by the Bowers-Liang model. However, our work is the first to obtain the URs of compact stars composed of interacting quark matter under the presence of anisotropy.

\section{Nonradial oscillations}

In addition to the URs already discussed above, it is also possible to provide empirical functions that do not depend on the specific EOS by involving the $f$-mode pulsation frequency \cite{Andersson1998, Lau2010, ChanSLL2014, Chirenti2015, ZhaoLattimer2022}. Such a mode is the lowest-order, fundamental, nonradial oscillation mode, characterized by a zero node number. This work also aims to study the correlation between the dimensionless frequency $\Omega_f= \tilde{M}\tilde{\omega}_f$ and other macroscopic properties of IQSs within an anisotropic context. The impact of anisotropic pressure on the $f$-mode frequency of compact stars using hadronic matter \cite{Curi2022, Mohanty:2023hha} and quark matter \cite{Curi2022, Arbanil:2023yil} has been recently investigated. In the case of QSs, the authors used the standard noninteracting quark matter EOS, i.e., the phenomenological MIT bag model EOS. We will extend such study considering strong interaction effects in dense matter of the compact star system. To do so, we will obtain the differential equations that govern the nonradial oscillations of anisotropic compact stars in the Cowling approximation for any anisotropy profile $\sigma$, as well as analyze in detail the different URs associated with the $f$-mode frequency. With this in mind, we start with the conservation equation for the energy-momentum tensor (\ref{EqEMTensor}), namely
\begin{align}\label{EqNablaT}
    \nabla_\mu T^{\mu\nu} =& \left[ (\nabla_\mu \rho)u^\mu + (\rho+ p_t)(\nabla_\mu u^\mu) \right]u^\nu + \mathcal{P}^{\mu\nu}\nabla_\mu p_t \nonumber  \\
    &+ (\rho+ p_t)a^\nu - \nabla_\mu(\sigma k^\mu k^\nu) = 0, 
\end{align}
where $a^\mu= u^\nu\nabla_\nu u^\mu$ is the four-acceleration of the fluid in the background metric and $\mathcal{P}^{\mu\nu}= g^{\mu\nu}+ u^\mu u^\nu$ is the projection tensor, which projects any tensor on the hypersurface orthogonal to $u_\mu$ \cite{RezzollaZanotti}. The nonradial oscillations equations are obtained by varying (\ref{EqNablaT}). Under the Cowling approximation \cite{Lindblom1990}, where the metric perturbations are neglected (namely, $\delta g^{\mu\nu}= 0$), the perturbation of the conservation equation $\delta(\nabla_\mu T^{\mu\nu})= 0$ implies that 
\begin{align}\label{EqNablaDelta}
    &\left[ (\delta u^\mu)\nabla_\mu\rho+ u^\mu\nabla_\mu\delta\rho+ (\delta\rho+ \delta p_t)(\nabla_\mu u^\mu)  \right.  \nonumber  \\
    &\left. + (\rho+ p_t)(\nabla_\mu\delta u^\mu) \right]u^\nu + \left[ u^\mu\nabla_\mu\rho+ (\rho+ p_t)(\nabla_\mu u^\mu) \right]\delta u^\nu \nonumber  \\
    & + (\nabla_\mu\delta p_t)u^\mu u^\nu + (\rho+ p_t)\left[(\delta u^\mu)\nabla_\mu u^\nu + u^\mu\nabla_\mu\delta u^\nu\right] \nonumber \\
    &+g^{\mu\nu}\nabla_\mu\delta p_t + (\delta u^\mu)u^\nu\nabla_\mu p_t + u^\mu(\delta u^\nu)\nabla_\mu p_t  \nonumber  \\
    &+ (\delta\rho+ \delta p_t)a^\nu - \nabla_\mu\delta(\sigma k^\mu k^\nu) = 0 ,
\end{align}
so that the projection of Eq.~(\ref{EqNablaDelta}) along the four-velocity, i.e., $u_\nu\nabla_\mu(\delta T^{\mu\nu})= 0$, leads to 
\begin{align}\label{ProAlongFourV}
    u^\mu\nabla_\mu\delta\rho &+ \nabla_\mu\left[ (\rho+ p_t)\delta u^\mu \right] + (\rho+ p_t)(\delta u^\nu)a_\nu  \nonumber  \\
    &+ u_\nu\nabla_\mu\delta(\sigma k^\mu k^\nu) =0, 
\end{align}
where we have used the normalization condition $u_\mu u^\mu= -1$ and the orthogonality condition $u_\mu a^\mu= 0$. Notice that using the property $u_\mu k^\mu =0$, Eq.~(\ref{ProAlongFourV}) can be written in the form given as in Ref.~\cite{Doneva2012}. Furthermore, projecting Eq.~(\ref{EqNablaDelta}) in the space orthogonal to the four-velocity, i.e., $\mathcal{P}^\alpha_{\ \nu}\nabla_\mu(\delta T^{\mu\nu})= 0$, we find 
\begin{align}\label{ProOrthFourV}
    &(\delta\rho+ \delta p_t)a^\alpha + \mathcal{P}^{\alpha\mu}\nabla_\mu\delta p_t + (\rho+ p_t)(\delta u^\mu)\nabla_\mu u^\alpha \nonumber  \\
    &+ (\rho+ p_t)u^\mu \nabla_\mu\delta u^\alpha - (\rho+ p_t)u^\alpha(\delta u^\nu)a_\nu  \nonumber  \\
    &- \mathcal{P}^\alpha_{\ \nu}\nabla_\mu\delta(\sigma k^\mu k^\nu) = 0 .
\end{align}

Fluid perturbation vectors inside a compact star can be decomposed in a basis of spherical harmonics. Thus, writing the Lagrangian displacement vector as \cite{Arbanil:2023yil}
\begin{equation}
\xi^{i}= \left(e^{-\Psi}W,-V\partial_{\theta},-\frac{V}{\sin^{2}\theta}\partial_{\phi}\right)\frac{Y_{\ell m}}{r^2} ,
\end{equation}
the explicit expression for the Eulerian variation of the four-velocity vector is given by
\begin{equation}
\delta u^{\mu}= \left(0,e^{-\Psi}\partial_{t}W,-\partial_{t}V\partial_{\theta},-\frac{\partial_{t}V}{\sin^{2}\theta}\partial_{\phi}\right)\frac{e^{-\Phi}Y_{\ell m}}{r^2} ,
\end{equation}
where $W=W(t,r)$ and $V=V(t,r)$ are fluid perturbation functions which depend on the time and radial coordinate, and $Y_{\ell m}=Y_{\ell m}(\theta,\phi)$ are the standard spherical harmonics. Taking into account the non-null components $k^1= e^{-\Psi}$ and $\delta k^0= e^{-2\Phi}\partial_{t}WY_{\ell m}/r^2$, Eq.~(\ref{ProAlongFourV}) provides the perturbation for energy density:
\begin{align}\label{EqDeltaRho}
    \delta\rho =& -(\rho+ p_r)\left[ \frac{e^{-\Psi}}{r^2}W'+ \frac{\ell(\ell+1)}{r^2}V \right]Y_{\ell m}  \nonumber  \\
    &- \rho'\frac{e^{-\Psi}}{r^2}WY_{\ell m} - 2\sigma\frac{e^{-\Psi}}{r^3}WY_{\ell m}  \nonumber  \\
    &- \sigma\frac{\ell(\ell+1)}{r^2}V Y_{\ell m} ,
\end{align}
where the prime indicates derivative with respect to $r$.

We assume a harmonic time dependence for the perturbation functions, namely $W(t,r)= W(r)e^{i\omega t}$, $V(t,r)= V(r)e^{i\omega t}$ and $\delta F(t,r,\theta,\phi)= \delta F(r)e^{i\omega t}Y_{\ell m}$, where $F$ represents any fluid variable and $\omega$ is the nonradial oscillation frequency to be determined. Consequently, the $\alpha= 1$ component of Eq.~(\ref{ProOrthFourV}) implies that 
\begin{equation}\label{EqAlpha1}
    (\delta\rho+ \delta p_r)\Phi' + (\delta p_r)'- \omega^2(\rho+ p_r)\frac{e^{\Psi- 2\Phi}}{r^2}W- \frac{2}{r}\delta\sigma = 0 ,
\end{equation}
while for the index $\alpha=2$, we obtain
\begin{equation}\label{EqAlpha2}
    \delta p_r+ \delta\sigma+ \omega^2(\rho+ p_r+ \sigma)e^{-2\Phi}V =0 .
\end{equation}

In view of Eqs.~(\ref{EqDeltaRho})-(\ref{EqAlpha2}) and taking into consideration that $\delta p_r= (dp_r/d\rho)\delta\rho$ due to the EOS $p_r= p_r(\rho)$, the nonradial perturbations of an anisotropic compact star in the Cowling approximation are described by 
\begin{widetext}
\begin{align}
    W' =&\ \frac{d\rho}{dp_r}\left[ (1+X)\omega^2r^2e^{\Psi-2\Phi}V + \Phi'W+ \frac{r^2e^{\Psi}\delta\sigma}{\rho+ p_r} \right] - \left( 1+\frac{d\rho}{dp_r} \right)\frac{2}{r}XW- \ell(\ell+1)(1+X)e^\Psi V ,  \label{nroEqW}  \\
    V' =& \left[ -\frac{\sigma'}{(\rho+p_r)(1+X)}+ 2\Phi' - \left( 1+\frac{d\rho}{dp_r} \right)\left( \Phi'+ \frac{2}{r}\right) \frac{X}{1+X} \right]V - \left( 1+\frac{d\rho}{dp_r} \right)\frac{e^{2\Phi}\Phi'\delta\sigma}{\omega^2(\rho+p_r)(1+X)}  \nonumber  \\
    &- \frac{e^\Psi}{r^2}\frac{W}{1+X} - \frac{e^{2\Phi}}{\omega^2(\rho+p_r)(1+X)}\left[ \frac{2}{r}\delta\sigma + (\delta\sigma)' \right] ,  \label{nroEqV}
\end{align}
\end{widetext}
with $X= \sigma/(\rho+p_r)$. It is important to remark that this system of differential equations is valid for any anisotropy profile $\sigma(r)$. It could be useful in future studies for the adoption of other functions $\sigma(r)$ that the literature provides. In particular, when the anisotropy vanishes (i.e., the quantities $\sigma$, $X$ and $\delta\sigma$ are zero) we recover the equations corresponding to isotropic stars \cite{Sotani2011}. In addition, for the anisotropy model (\ref{QLProfileEq}), that is, $\sigma= \beta_{\rm QL}\mu p_r$, we have $\delta\sigma= (\partial\sigma/\partial p_r)\delta p_r$. This is because the variation $\delta\mu$ is associated with the variation of the metric potential $\Psi$, which is zero in the Cowling approximation. In this case, the perturbation for the anisotropy factor can be written as
\begin{equation}\label{EqDeltaSigma}
    \delta\sigma= -\frac{\partial\sigma}{\partial p_r}\left( 1+\frac{\partial\sigma}{\partial p_r} \right)^{-1}\omega^2(\rho+p_r)(1+X)e^{-2\Phi}V. 
\end{equation}

By substituting Eq.~(\ref{EqDeltaSigma}) into the general system of time-independent differential equations (\ref{nroEqW}) and (\ref{nroEqV}), we hence manage to retrieve the nonradial oscillation equations for the QL model, namely \cite{Arbanil:2023yil, Doneva2012}
\begin{align}
    W' =&\ \frac{d\rho}{dp_r}\left[ \left(1+ X\right)\left(1+\frac{\partial\sigma}{\partial p_r}\right)^{-1}\frac{\omega^2r^2V}{e^{2\Phi-\Psi}}+\Phi'W \right] \nonumber  \\
    &- X\left[ \left(1+\frac{d\rho}{dp_r}\right)\frac{2W}{r}+\ell(\ell+1)e^{\Psi}V \right]  \nonumber \\
    &- \ell(\ell+1)e^{\Psi}V ,  \label{nroQLEqW} 
\end{align}
\begin{align}
    V' =&\ V\left[ -\frac{\sigma'}{\rho+p_r+\sigma}-\left(1+\frac{d\rho}{dp_r}\right)\left(\Phi'+\frac{2}{r}\right)\frac{X}{1+X} \right.  \nonumber \\
    &\left.+ \frac{2}{r}\frac{\partial\sigma}{\partial p_r}+\left(1+\frac{\partial\sigma}{\partial p_r}\right)^{-1}\left(\frac{\partial^2\sigma}{\partial p_r^2}p_r'+\frac{\partial^2\sigma}{\partial\mu\partial p_r}\mu'\right)\right]  \nonumber \\
    &+ 2\Phi'V-\left(1+\frac{\partial\sigma}{\partial p_r}\right)\frac{e^{\Psi}W}{r^2(1+X)} . \label{nroQLEqV} 
\end{align}

\begin{table*}
\caption{\label{table4} Fitting coefficients for the empirical formulas of the $f-\Lambda$, $f-\bar{I}$ and $f-C$ relations given in Eqs.~(\ref{UREqOmefLambda}), (\ref{UREqOmefI}) and (\ref{UREqOmefC}), respectively. The numerical values for the reduced chi-squared $(\chi_{\rm red}^2)$ are shown in the last row. }
\begin{ruledtabular}
\begin{tabular}{cccc|cccc|cccc}
\multicolumn{4}{c}{$f-\Lambda$}  &  \multicolumn{4}{c}{$f-\bar{I}$}  &  \multicolumn{4}{c}{$f-C$}  \\
\hline
$\beta_{\rm QL}$  &  $-0.6$  &  0  &  0.6  &  $\beta_{\rm QL}$  &  $-0.6$  &  0  &  0.6  &  $\beta_{\rm QL}$  &  $-0.6$  &  0  &  0.6  \\
\hline
  $d_0 [10^{-1}]$  &  1.6688  &  2.0329  &  2.2532  &  $g_0 [10^{-3}]$  &  $-3.8155$  &  $-7.2608$  &  $-7.0509$  &  $h_0 [10^{-3}]$  &  $-2.0622$  &  $-1.5756$  &  $-3.9459$  \\
  $d_1 [10^{-2}]$  &  2.1977  &  $-0.2391$  &  $-1.7145$  &  $g_1 [10^{-1}]$  &  $2.6599$  &  $4.6907$  &  $4.0156$  &  $h_1 [10^{-1}]$  &  $2.4545$  &  $2.1832$  &  $3.1808$  \\
  $d_2 [10^{-2}]$  &  $-3.8001$  &  $-2.9844$  &  $-2.1979$  &  $g_2$  &  $-1.3375$  &  $-5.1309$  &  $-3.0898$  &  $h_2$  &  $2.2344$  &  $2.4945$  &  $1.5897$  \\
  $d_3 [10^{-2}]$  &  $1.0384$  &  $0.8604$  &  $0.5514$  &  $g_3 [10^1]$  &  $1.9499$  &  $4.8932$  &  $2.9761$  &  $h_3$  &  $-4.1448$  &  $-3.9546$  &  $-0.1437$  \\
  $d_4 [10^{-3}]$  &  $-1.3161$  &  $-1.0811$  &  $-0.4146$  &  $g_4 [10^2]$  &  $-0.8627$  &  $-1.9168$  &  $-1.0978$  &  $h_4$  &  $-0.5363$  &  $-0.8226$  &  $-5.3326$  \\
  $d_5 [10^{-5}]$  &  $8.3270$  &  $6.7439$  &  $-0.1876$  &  $g_5 [10^2]$  &  $1.7164$  &  $3.5051$  &  $1.9141$  &  $--$  &  $--$  &  $--$  &  $--$  \\
  $d_6 [10^{-6}]$  &  $-2.1389$  &  $-1.7410$  &  $0.9861$  &  $g_6 [10^2]$  &  $-1.3045$  &  $-2.4521$  &  $-1.2975$  &  $--$  &  $--$  &  $--$  &  $--$  \\

  $\chi_{\rm red}^2 [10^{-6}]$  &  $0.9891$  &  $8.9843$  &  $7.1620$  &  $\chi_{\rm red}^2 [10^{-6}]$  &  $1.5686$  &  $9.6502$  &  $7.8122$  &  $\chi_{\rm red}^2 [10^{-6}]$  &  $0.2385$  &  $5.0862$  &  $0.2412$  \\
\end{tabular}
\end{ruledtabular}
\end{table*}

The above equations (\ref{nroQLEqW}) and (\ref{nroQLEqV}) will be solved numerically from the center up to the stellar surface of the anisotropic sphere. For this purpose, suitable boundary conditions have to be established. Thus, at $r=0$, we assume that the radial variables $W$ and $V$ take the respective forms
\begin{align}\label{CenCond_NRO}
W &= cr^{\ell+1},  &  V &= -c\frac{r^{\ell}}{\ell},
\end{align}
where $c$ stands for a dimensionless constant. Meanwhile, at $r= R$, the following condition has to be satisfied.
\begin{equation}\label{SurCond_NRO}
\left(1+X\right)\frac{\omega^2V}{e^{2\Phi}} + \left[1+\frac{\partial\sigma}{\partial p_r}\right]\left[\frac{r\Phi'}{2}-X\right]\frac{2W}{e^{\Psi}r^3} =0 ,
\end{equation}
and in our calculations we will deal with the quadripolar modes, that is, when $\ell =2$. Note further that the dimensionless rescaled form for the new variables involved in the nonradial perturbations is given by
\begin{align}
    \tilde{\omega} &= \frac{\omega}{\sqrt{4B_{\rm eff}}},  &  \tilde{W} &=  (4B_{\rm eff})^{3/2}W,  &  \tilde{V} &=  (4B_{\rm eff})V .
\end{align}

\begin{figure*}
 \includegraphics[width=17.8cm]{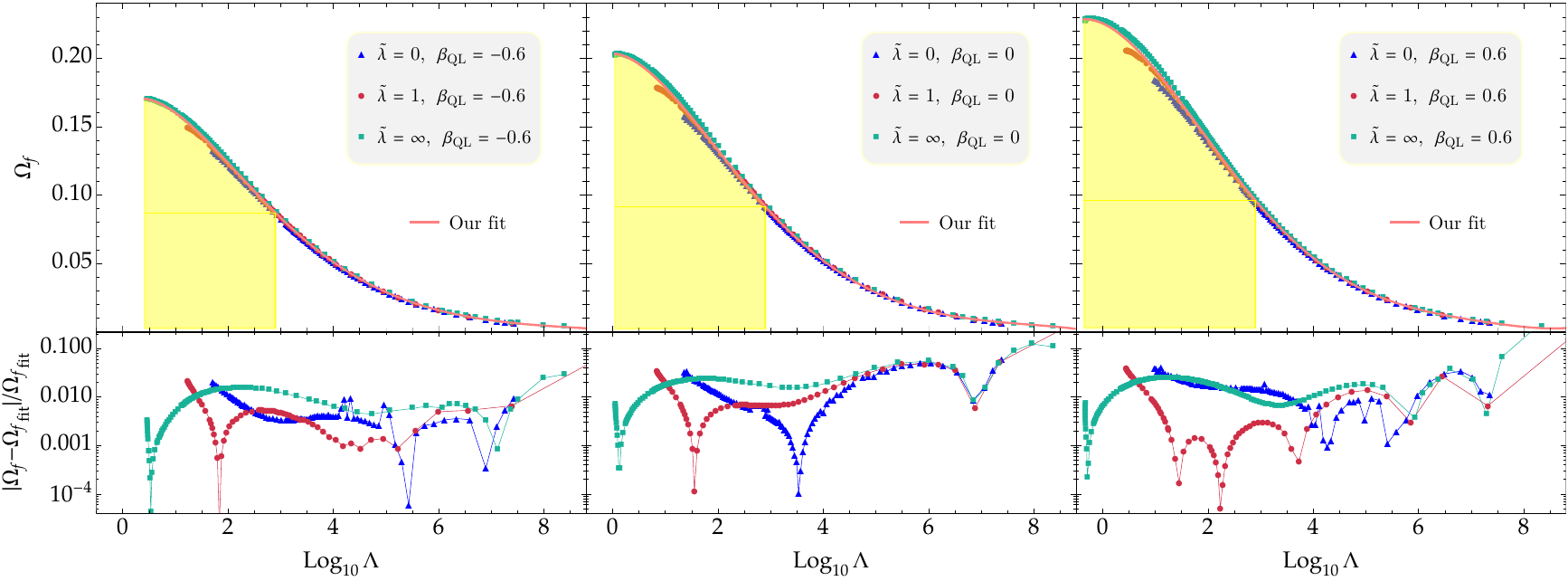}
  \includegraphics[width=17.8cm]{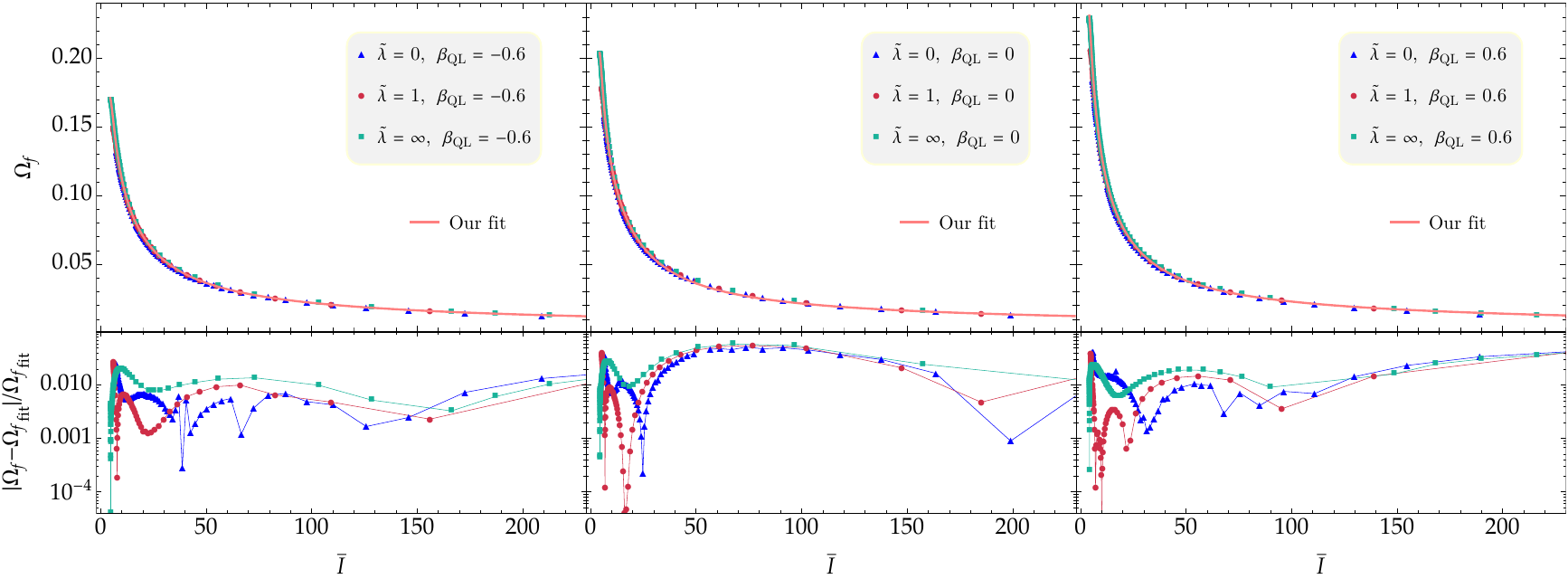}
   \includegraphics[width=17.8cm]{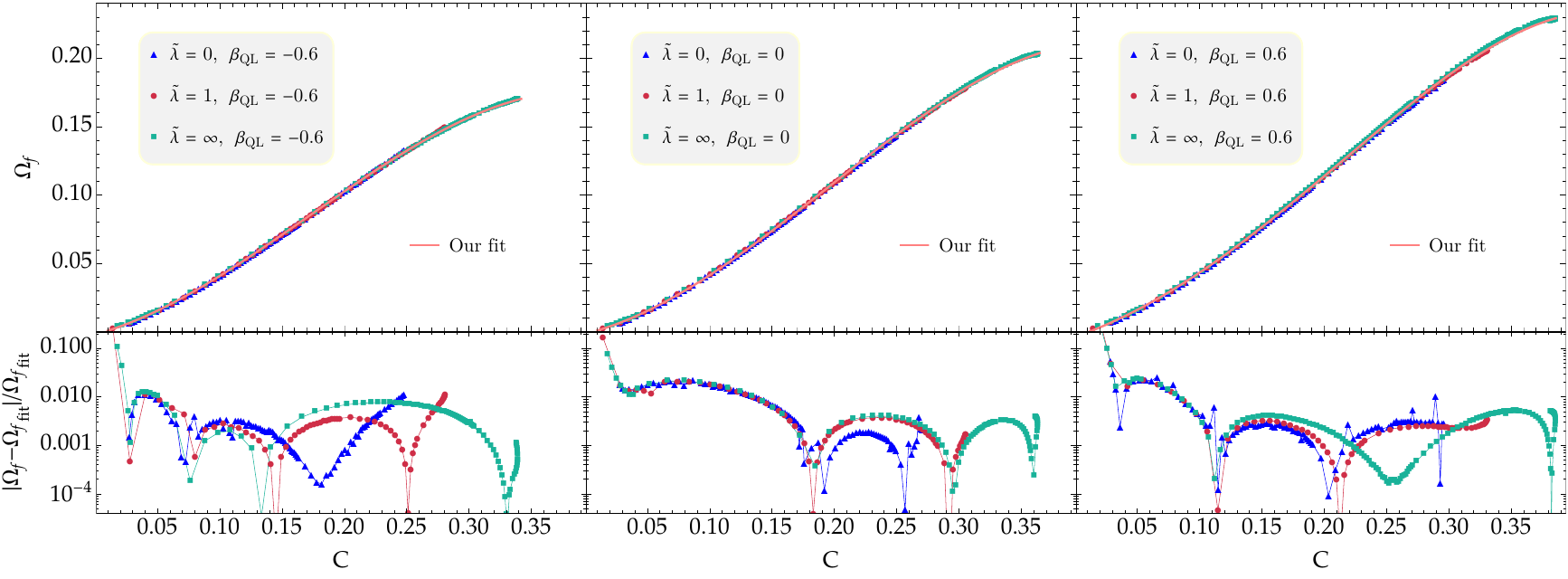}
 \caption{\label{figOLUnivRelationQL} From top to bottom, dimensionless $f$-mode frequency $\Omega_f$ as a function of (a) the tidal deformability $\Lambda$, (b) the moment of inertia $\bar{I}$, (c) the compactness $C$ for anisotropic IQSs. The color and symbol conventions follow those of Fig.~\ref{figUnivRelationsQL}, where the pink curves represent our fitting functions Eq.~(\ref{UREqOmefLambda}), Eq.~(\ref{UREqOmefI}), Eq.~(\ref{UREqOmefC}) of corresponding URs, with expansion coefficients listed explicitly in Table \ref{table4}. The lower panel in each plot displays the relative fractional errors between fits and the numerical results. As in Fig.~\ref{figUnivRelationsQL}, the yellow area in the first row represents the EOS-independent constraint $\Lambda_{1.4} \leq 800$ from Ref.~\cite{Abbott2017PRL}. }  
\end{figure*}

Chan et al.~\cite{ChanSLL2014} found EOS-insensitive empirical formulas which relate the $f$-mode frequency and the tidal deformability of compact stars, so here we focus on this type of universal correlation and refer to it as the $f-\Lambda$ relation. Given a $\rho_c$ for a specific EOS and a value of $\beta_{\rm QL}$, we solve the system of equations (\ref{nroQLEqW}) and (\ref{nroQLEqV}) from the center up to the stellar surface for a set of test values $\omega^2$ satisfying the condition (\ref{CenCond_NRO}) at $r=0$. The appropriate frequencies will then be obtained when the boundary condition at the surface (\ref{SurCond_NRO}) is obeyed. This procedure is carried out for a range of central densities, i.e., for all the stellar configurations shown in Fig.~\ref{fig_MR_I_L}. As we already mentioned, we are interested in calculating the fundamental mode frequency $\omega_f$. Thus, the dimensionless $f$-mode oscillation frequencies $\Omega_f$ as a function of the tidal deformability $\Lambda$ are presented in Fig.~\ref{figOLUnivRelationQL}a for several values of the parameters $\tilde{\lambda}$ and $\beta_{\rm QL}$. We find the following empirical expression
\be\label{UREqOmefLambda}
  \Omega_f = \sum_{n=0}^6 d_n \left( \log_{10}\Lambda \right)^n ,
\ee
which is represented by the pink curve and the fitting coefficients $d_n$ with their corresponding reduced chi-squared ($\chi^2_{\rm red}$) values are given in Table \ref{table4}. We observe that the $\chi^2_{\rm red}$ value gets largely (slightly) reduced when adding negative (positive) anisotropy, indicating that anisotropy can enhance this UR.

It was shown that the $f$-mode also has a connection with the moment of inertia \cite{Lau2010}. In that regard, in Fig.~\ref{figOLUnivRelationQL}b we analyze the correlation between the $f$-mode frequency $\Omega_f$ and the normalized moment of inertia $\bar{I}$, where the numerical data are fitted by the following function
\be\label{UREqOmefI}
  \Omega_f = \sum_{n=0}^6 g_n \left( \frac{1}{\bar{I}} \right)^{n/2} .
\ee
The fitting parameters $g_n$ are listed in Table \ref{table4}.
For the isotropic case (i.e., when $\beta_{\rm QL}= 0$ in the middle panel), we see that this $f-\bar{I}$ relation is insensitive to variations of $\tilde{\lambda}$ with maximum deviations at the $6\%$ level. Nonetheless, this fractional difference decreases slightly when we introduce anisotropy within the IQS, see the specific cases for $\beta_{\rm QL}= -0.6$ and $0.6$ in the side panels.  This trend also manifests in the $\chi^2_{\rm red}$ values given in Table \ref{table4}.

Finally, we will examine the relation between the $f$-mode frequency and the compactness of anisotropic IQSs. Such a $f-C$ relation is illustrated in Fig.~\ref{figOLUnivRelationQL}c, where the pink curve represents the power series expansion
\be\label{UREqOmefC}
  \Omega_f = \sum_{n=0}^4 h_n C^n ,
\ee
with fitting coefficient $h_n$ given in Table \ref{table4}. Remarkably, for the $f-C$ UR as well as for the other correlations involving the $f$-mode, the presence of anisotropic pressure (for both positive and negative $\beta_{\rm QL}$) decreases the value of $\chi_{\rm red}^2$ with respect to the isotropic scenario. This means that these EOS-independent universal relations become stronger when there is anisotropy within the IQS.

Similar to the other universal relations involving $\Lambda$, we can also obtain a bound for the fundamental mode frequency from the tidal deformability constraint $\Lambda_{1.4} \leq 800$ \cite{Abbott2017PRL}. Thus, the $f-\Lambda$ relation (see yellow zone in the middle top panel of Fig.~\ref{figOLUnivRelationQL}) leads to the canonical normalized $f$-mode frequency $\Omega_{f,1.4} \geq 0.091$ for the isotropic scenario. Taking into consideration that $\Omega_f= GM\omega_f/c^3$ in physical units, this inferred result is equivalent to $f_{f,1.4} \geq 2.108\, \rm kHz$. Meanwhile, the limits for $f_{f,1.4}$ within the anisotropic context have been reported in the last column of Table \ref{tableFromGW170817}.

\section{Summary and final remarks}\label{conclusion}
In this study, we have systematically explored the various $I-\text{Love}-C-f$ universal relations of anisotropic IQSs, with QL anisotropy profile and a general interacting quark matter EOS that subsumes color superconductivity effect, pQCD corrections, bag constant and different phases in a single parameter $\tilde{\lambda}$. We have shown that $I-\text{Love}-C-f$ universal relations generally hold, with maximum deviations to the polynomial fit at or below percent levels when varying QM EOS parameter $\tilde{\lambda}$ and anisotropy parameter $\beta_{\rm QL}$. With analyses of the reduced $\chi_{\rm red}^2$ values, we find that more positive anisotropy tends to enhance the $\bar{I}-\Lambda$ and $\bar{I}-C$ URs, but weakens the $C-\Lambda$ UR. For all the URs involving $f$-mode frequency, we find that they are enhanced by the inclusion of anisotropy (whether positive or negative).  Furthermore, by means of these URs and from the tidal deformability constraint $\Lambda_{1.4} \leq 800$, given by the GW170817 event \cite{Abbott2017PRL}, we have been able to impose limits on the various canonical macroscopic properties corresponding to a $1.4M_\odot$ compact star.

Note that in the calculations of nonradial oscillations, we heve adopted Cowling approximations, where the typical error in the isotropic case is around $20\%$ percent level compared to the full GR treatment~\cite{ZhaoLattimer2022}.  Full GR examinations of nonradial oscillations in the anisotropic case is a topic that was rarely explored even for neutron stars~\cite{mondal2023non}. We leave such full GR treatments for future studies.

Besides, in addition to the quasi-local anisotropy ansatz, it is also worthwhile to test other anisotropy models. Moreover, strange quark matter may be localized as solid clusters called strangeons~\cite{Xu:2003xe,Miao:2020cqj,Lai:2022yky}, which can form strangeon stars that are also highly compact~\cite{Lai:2009cn,Lai:2017ney,Gao:2021uus,Li:2022qql,Chen:2023pew,Zhang:2023mzb,Zhang:2023szb}. Recently, one of the authors proposed the new possibility of inverted hybrid stars~\cite{Zhang:2022pse,Zhang:2023zth,Negreiros:2024cvr}. The anisotropic effect on these new types of compact stars and related universal relations may also manifest interesting features. We leave these for future studies.

\begin{acknowledgments}
We thank Dr.~Zhiqiang Miao for the helpful discussions regarding detection aspects on $f$-mode. JMZP acknowledges support from ``Fundação Carlos Chagas Filho de Amparo à Pesquisa do Estado do Rio de Janeiro'' -- FAPERJ, Process SEI-260003/000308/2024. C. Zhang is supported by the Jockey Club Institute for Advanced Study at The Hong Kong University of Science and Technology. 
\end{acknowledgments}\


%

\end{document}